\documentstyle[11pt,aaspp4]{article}

\begin{document}

\title{
	Near Infrared Imaging of the Hubble Deep Field
	with The Keck Telescope\altaffilmark{1}
}

\author{
	David W. Hogg\altaffilmark{2},
	G. Neugebauer\altaffilmark{3},
	Lee Armus\altaffilmark{3},
	K. Matthews\altaffilmark{3},
	Michael A. Pahre\altaffilmark{4},
	B. T. Soifer\altaffilmark{3} and
	A. J. Weinberger\altaffilmark{3}
}

\altaffiltext{1}{
Based on observations obtained at the W. M. Keck Observatory, which is
operated jointly by the California Institute of Technology and the
University of California; and on observations obtained with the
NASA/ESA Hubble Space Telescope, which is operated by AURA under NASA
contract NAS~5-26555.}
\altaffiltext{2}{
Theoretical Astrophysics, California Institute of Technology, Mail
Stop 130-33, Pasadena CA 91125; {\tt hogg@tapir.caltech.edu}}
\altaffiltext{3}{
Palomar Observatory, California Institute of Technology, Mail Stop
320-47, Pasadena CA 91125; {\tt gxn}, {\tt lee}, {\tt kym}, {\tt bts}
and {\tt alycia@mop.caltech.edu}}
\altaffiltext{4}{
Palomar Observatory, California Institute of Technology, Mail
Stop 105-24, Pasadena CA 91125; {\tt map@astro.caltech.edu}}

\begin{abstract}
Two deep $K$-band ($2.2~{\rm \mu m}$) images, with point-source
detection limits of $K=25.2$~mag (one sigma), taken with the Keck
Telescope in subfields of the Hubble Deep Field, are presented and
analyzed.  A sample of objects to $K=24$~mag is constructed and
$V_{606}-I_{814}$ and $I_{814}-K$ colors are measured.  By stacking
visually selected objects, mean $I_{814}-K$ colors can be measured to
very faint levels; the mean $I_{814}-K$ color is constant with
apparent magnitude down to $V_{606}=28$~mag.
\end{abstract}


% -----------------------------------------------------------------------------
\section{Introduction}

The Hubble Deep Field (HDF; Williams et al 1996) is an imaging project
with the Hubble Space Telescope (HST) in a single field in four
bandpasses: $U_{300}$, $B_{450}$, $V_{606}$ and $I_{814}$.  The HDF
images are the deepest images of the sky ever taken; roughly two
million sources can be found per square degree to $V_{606}\approx
30$~mag.  Near infrared data in the HDF are essential for
understanding the significance of old stellar populations in faint
objects, measuring the visual radiation from objects at high redshift,
and finding or placing limits on extremely red objects.  Until the
servicing mission planned for 1997, HST has no near-infrared imaging
capability; even after the NICMOS instrument is installed, it will
have limited sensitivity longward of $2~{\rm \mu m}$.  For these
reasons, we have taken two very deep $K$-band ($2.2~{\rm \mu m}$)
images of subfields of the HDF with the Keck Telescope as part of a
near infrared survey.  In this paper we present a
near-infrared--selected sample, color distributions for samples
selected visually and in the near infrared, and color-morphology
relations.  Although this near infrared survey is still underway, we
are releasing these data to the community via the
internet\footnote{\tt http://www.cco.caltech.edu/\~{}btsoifer/hdf.html}.

% -----------------------------------------------------------------------------
\section{Observations}

Two subfields in the HDF, the first (NIRC-HDF-A) centered on the
HST/WF4 image, and the second (NIRC-HDF-B) centered on the HST/PC1
image, as shown in Fig.~1, were observed with the Near Infrared Camera
(NIRC; Matthews \& Soifer 1994) on the Keck Telescope.  The camera
sits at the $f/25$ forward Cassegrain focus of the telescope and
contains a $256\times 256$ Santa Barbara Research Center InSb detector
with a scale of $0.15~{\rm arcsec~pixel^{-1}}$.  The instantaneous
field of view is $\approx 38\times 38~{\rm arcsec^2}$.

The observations were taken with the $K$-band filter
($2.00\leq\lambda\leq 2.43~{\rm \mu m}$) on 4 and 5 February 1996 UT
for 29,760 sec (NIRC-HDF-A) and on 6 and 7 February 1996 UT for 27,000
sec (NIRC-HDF-B).  The air mass of the observations ranged from 1.40
to 1.75; 68~percent had an air mass $\leq 1.50$. The conditions were
clear and apparently photometric, although there was some variation in
the sky brightness.  The stellar source in the NIRC-HDF-A field has a
full width at half maximum (FWHM) of $0.75$~arcsec in the final,
processed images. The observations were taken in sets of nine two
minute exposures, each set offset so that an object in the sky fell at
the center of, at the corners of and at the midpoints of the sides of
a square 10~arcsec on a side.  The pointing of the center of each set
of nine exposures was jogged in an arbitrary direction by about
0.75~arcsec.  Each two minute exposure consisted of eight 15~s
integrations coadded into a single image. An offset guider autoguiding
on a visual image maintained the tracking of the telescope. The
NIRC-HDF-B observations were interrupted by an earthquake and a
failure in the telescope active control system both of which required
an interruption of about an hour each to re-collimate the telescope;
other than on these occasions, the measurements were contiguous during
each night's observing. The photometric sensitivity was determined by
measuring faint standard stars of Persson (1996) and Casali \&
Hawarden (1992).  The uncertainty in the photometric calibration is
less than $0.05$~mag.

The data were processed by creating a sky image for each set of nine
images to eliminate artifacts introduced by the field rotation
resulting from the alt--azimuth configuration of the Keck Telescope.
Bright objects, identified by iterative processing, were masked to
produce these sky images for background subtraction and flatfielding.
Integer-pixel registration to combine the individual images was
accomplished by finding the centroid of the brightest object in the
image.  This object was visible in essentially all the frames; in
those few cases where it was off the frame, its coordinates were
transferred from a secondary object which was bright enough to appear
in individual images.  The individual registered images were averaged
to produce the final images shown in Figs.~2 and 3.  The
pixel-to-pixel rms values in the central parts of the images
correspond to $25.4$ and $25.2$~mag in $1~{\rm arcsec^2}$ in the
NIRC-HDF-A and NIRC-HDF-B images, respectively.  Because an aperture
of diameter $1.5\times{\rm FWHM}$ encloses a solid angle of $1~{\rm
arcsec^2}$, these are also roughly the one-sigma point-source
detection limits.

For visual data the ``version 2'' reductions and calibrations of the
HST/WFPC2 images of the HDF (Williams et al 1996) in the F814W
($I_{814}$) and F606W ($V_{606}$) bands were used.

% -----------------------------------------------------------------------------
\section{Analysis}

\subsection{Source detection}

To construct a $K$-selected sample, the images in the $K$-band mosaics
were convolved with a Gaussian with the same FWHM as the seeing
($0.75$~arcsec).  The convolved image was divided by a non-uniform rms
(root variance) image whose value at each pixel is proportional to the
reciprocal square root of the number of images contributing to that
pixel.  Sources more significant than $2\,\sigma$ at their centers in
this image are included in the $K$-selected sample, where here
$\sigma$ is the rms noise in this convolved image.  Because the
sources are smaller than the angular extent over which the rms varies
significantly in the original mosaics, the sample constructed this way
is significance-limited.  However, the sample is not uniformly
flux-limited because the rms does vary.  A large fraction of spurious
sources were removed by requiring that the flux through a $1.5$~arcsec
diameter aperture centered on the detected source be brighter than
$K=24.2$~mag in the original, unconvolved mosaic.  By this procedure,
there are 44 $K$-selected sources in NIRC-HDF-A and also 44 in
NIRC-HDF-B.  The sources are listed in Tables~1 and 2.

The completeness of the sample can be estimated by comparison with
$K$-band galaxy counts in other blank fields (Djorgovski et al 1995).
This comparison indicates that the completeness is above 90~percent to
$K=23$~mag and falls off rapidly to $K=24$~mag.  These numbers are
perhaps slightly worse than the completeness levels of similar data
(Djorgovski et al 1995).  The data presented here have slightly lower
sensitivity, probably because they were taken in conditions of more
rapidly changing sky brightness.  Furthermore, the detection
procedures differ.  The contamination by spurious (i.e., noise)
sources was estimated by running the same detection algorithm
(including the cut at $K=24.1$~mag) but searching for negative instead
of positive sources.  Only three negative sources were found in the
two images, suggesting that spurious source contamination is only a
few percent.  Of course this test does not correct for spurious
sources due to noise fluctuations which happen to lie in the outer
profiles of real objects, so it is not a truly conservative estimate
of spurious source contamination.

For a visually selected sample, the HDF team ``version 2'' catalog, in
which sources were detected in a summed $(V_{606}+I_{814})$-band image
by Williams et al (1996), was employed.  The AB magnitudes used by the
HDF team have been converted to magnitudes zero-pointed to Vega for
this work.  The conversions are $V_{606}=V_{606,AB}-0.12$~mag and
$I_{814}=I_{814,AB}-0.44$~mag.  For morphological type information we
employed the ``RSE'' classification scheme presented in the
van~den~Bergh et al (1996) catalog.

\subsection{Photometry}

Both isophotal and aperture $K$-band magnitudes were measured for all
$K$-selected sources.  Isophotal magnitudes were measured out to a
surface brightness of $K=22.6$~mag in $1~{\rm arcsec^2}$.  Aperture
magnitudes were measured in $1.5$~arcsec diameter apertures in the
$K$-band images; roughly twice the seeing FWHM.  A correction of
$\Delta K=-0.10$~mag was added to the aperture magnitudes to account
for flux outside the aperture, under the assumption that the sources
are roughly point-like.  A ``total'' $K$-band magnitude was assigned
to each source by taking the brighter of the isophotal or corrected
aperture magnitudes.  These total $K$-band magnitudes are given in
Tables~1 and 2.  The uncertainties include the uncertainty in the flux
measurement and the uncertainty in the background level estimate,
measured individually for each aperture magnitude in an annulus from
2.25 to 3.38~arcsec diameter, and for each isophotal magnitude in a
strip of width 3.6~arcsec surrounding the outer isophote.

\subsection{Colors}

In order to measure colors, the $V_{606}$ and $I_{814}$-band images
were smoothed with the Gaussian kernel which makes the FWHM of stars
in the smoothed and $K$-band images agree.  Fluxes were measured in
the smoothed and $K$-band images through $1.5$~arcsec diameter
apertures.  These aperture magnitudes were subtracted to make
$V_{606}-I_{814}$ and $I_{814}-K$ colors.  Colors were measured
through $1.5$~arcsec diameter apertures even if the total magnitude
was obtained from an isophotal magnitude.  Colors thus derived are
given in Tables~1 and 2, and a color-magnitude diagram for the
$K$-selected sample is shown in Fig.~4.  The uncertainties in the
colors include the uncertainties in both flux measurements, and in
both sky estimations.  Not all sources found in NIRC-HDF-B are shown
in Fig.~4 because some do not lie within the HST image.  In the
overlap region, however, all sources detected in the $K$-band images
are apparent in the $I_{814}$-band except for a few marginal
candidates which are consistent with being spurious sources in the
$K$-band image.  The color-color diagram is shown in Fig.~5.

Some visually selected sources are not apparent in the $K$-band
images.  However, the great depth of the HST images allows a study of
the mean $K$-band fluxes of sources below the $K$-band detection limit
by averaging together large numbers of distinct visually selected
sources.  Small image sections from the smoothed-$V_{606}$,
smoothed-$I_{814}$ and $K$-band images were cut out around all sources
in the one-magnitude-wide bins centered on each whole and half
magnitude value in the $V_{606}$-band.  These image sections were
averaged together to make ``mean'' images, one in each band, for each
magnitude bin.  Only objects in the central parts of the $K$-band
images, where the noise is lowest, were used for constructing the mean
galaxies, and sources within 2~arcsec of a $K<19$~mag source were
excluded.  In these mean images, the mean source is clearly apparent
in all three of the images ($V_{606}$, $I_{814}$ and $K$), standing
out at the center above the mean background level.  Tests in which
random image locations are stacked showed no mean sources; the tests
show only a mean background level.  A mean color for each magnitude
bin was measured from these mean images by exactly the same procedure
as was used for the individual source photometry.  The results are
shown in Fig.~6.  Because the uncertainties are dominated by scatter
in the population rather than by photometric measurement uncertainty,
Fig.~6 shows root-mean-square uncertainties calculated by
bootstrap-resampling of the data (Efron \& Tibshirani, 1991).  The
mean $I_{814}-K$ colors are fairly constant with magnitude down to the
practical limit of this technique, $V_{606}=28$~mag.  It is worthy of
note that these are the faintest levels at which visual--near-infrared
colors have ever been reliably measured.

\subsection{Morphology}

To exploit the resolution of the HST images, morphological parameters
were measured in the $I_{814}$-band images.  These parameters include
radii, ellipticities and orientations.  No strong correlation with
$K$-band magnitude or $I_{814}-K$ color was found for any of these
quantities.  In addition, the $K$-band images can be compared with the
Gaussian-smoothed $I_{814}$-band images (prepared for the purpose of
color measurement; see above) in order to look for any
wavelength-dependence of morphology.  There is no evidence for
differences in morphologies between the $I_{814}$ and $K$-band images
of the sources, i.e., no apparent differences in extent, ellipticity,
orientation, or central concentration.

A catalog of HDF sources with morphological classification has been
constructed by van~den~Bergh et al (1996) including several different
classification schemes.  We make use of their ``RSE'' scheme, which
sorts sources into elliptical (E), spiral (S) and peculiar (P)
categories.  This sample excludes the PC1 data and therefore has
little overlap with NIRC-HDF-B.  It also excludes the two brightest
objects in the NIRC-HDF-A region of WF4; these are brighter than the
$21<I_{814}<25$~mag range which was classified.  Colors were measured
for all objects with morphological classifications in NIRC-HDF-A and
they are shown in Fig.~7.

% -----------------------------------------------------------------------------
\section{Discussion}

The very blue object in NIRC-HDF-A with $K=20.8$~mag (at
$\Delta\alpha,\Delta\delta=9,-1$ in Fig.~2) can be identified with the
stellar object at the center of the WF4 image.  This is presumed to be
a Galactic star but it has not been spectroscopically identified.
Such identification will occur when spectroscopic surveys in the field
are completed.  Indicated on Figs.~2 and 3 are the four very red
sources, with $I_{814}-K>4$~mag, the expected color range for old
stellar populations at $z>1$ (Aaronson 1977; Coleman, Wu \& Weedman
1982).  Also marked are those sources with redshifts from two
spectroscopic surveys (Steidel et al 1996; Cohen et al 1996).

Although we have made no concerted effort to construct source number
counts, which would involve accounting for varying exposure times over
the images and incompleteness at the faint end, the numbers of sources
are consistent with previous deep infrared counts based on similar
data (Djorgovski et al 1995).  In the central (i.e., low-noise) parts
of the images, there are about 70 beams per source, taking a beam to
be an aperture of diameter $1.5$ times the seeing FWHM.  Because no
correction for incompleteness has been applied, 70 is an upper limit.
If the counts continue to increase by a factor of roughly two per
magnitude, as they do for $K<24$~mag (Djorgovski et al 1995), and an
estimated correction for completeness is applied, the practical
confusion limit of about 25 beams per source ought to be reached at
$K=25$~mag at this seeing.

Except for the few very red sources, the source colors (Figs.~4 and 5)
are consistent with both old stellar populations observed in the
redshift range $0<z<1$ and younger populations observed at any
redshift $<5$ (Aaronson 1977; Coleman et al 1980).  There is no
$K<23.5$~mag source in the $K$-band images which is not apparent in
the visual images, and there is no $V<25.5$~mag visual source which
does not have positive flux (at the $1\,\sigma$ level at least) in the
low-noise regions of the $K$-band images.

The fact that the mean $I_{814}-K$ colors shown in Fig.~6 are
relatively constant with magnitude implies, under fairly robust
assumptions, that the count slope in the $K$ band over the magnitude
range $22<K<25.5$~mag ought to be very similar to that in the
$I_{814}$ band, about $0.18<d\log N/dm<0.31$ (Williams et al 1996).
This result is not consistent with faint galaxy models in which, on
average, apparently fainter objects are older.  In particular, we rule
out the strong upturn in the faint $K$-band galaxy counts predicted in
``fading dwarf'' models of the faint galaxy counts (Babul \& Ferguson
1996).  Of course, if there is a population of fading dwarfs with very
low surface brightnesses, or a population which merges into larger
objects as they fade, an upturn in $K$-band counts or $I-K$ colors
might be avoided.

The lack of apparent differences in morphology between the $I_{814}$
and $K$ bands suggests that there are not large color gradients in
faint optical sources, at least at the angular resolution of these
$K$-band images.  This may argue against some models which make the
faint galaxies small starburst knots inside the potential wells of
large, forming galaxies at $z>1.5$ (Katz 1992; Pascarelle et al 1996),
which might show extended halos of (slightly) older stars.  Since the
color-selected high-redshift sources of Steidel et al (1996) are good
candidates for being young starforming knots in primeval galaxies, it
is particularly interesting that even they show no strong color
gradients from the rest-frame far ultraviolet to the rest-frame
visual.

The detailed morphologies of local, bright galaxies change
dramatically between the visual and ultraviolet (Giavalisco et al
1996; O'Connell \& Marcum 1996; Abraham 1996) which would be observed
as the near-infrared and visual, respectively, at high redshift.  The
lack of color gradients may suggest, therefore, that these faint
sources are not ``grand-design'' galaxies at intermediate or high
redshift.  This may argue for models in which the bulk of faint
sources are intrinsically small, low-luminosity galaxies at modest
redshifts, but this will remain uncertain until their redshifts or
physical properties can be constrained in other ways.

As is clear from Fig.~7, there is little or no correlation between the
visual--near-infrared color and the morphological type assigned by
van~den~Bergh et al (1996).  This may be caused by the assignment of
``E'' classification to blue sources with small angular sizes, which
may not be classical ellipticals, or it may cast doubt upon the
usefulness of classical morphological studies in the observed visual
band at very faint levels or very high redshifts.

The two $K$-band images presented here represent the first results
from a deep near-infrared survey in the HDF.  At least two more
pointings to comparable depth in the HDF, along with some $L$-band
imaging ($3.5~{\rm \mu m}$), are planned.  Near-infrared images of the
HDF will be acquired when the NICMOS instrument is installed in HST.
The superb spatial resolution and sensitivity of NICMOS in the $J$ and
$H$ bands will make such observations extremely valuable.  On the
other hand, the higher sensitivity of Keck/NIRC in the $K$ band makes
the groundbased observations presented here complementary.  We look
forward to the impact of imaging at high angular resolution, all the
way to $3.5~{\rm \mu m}$, when the adaptive optics system has been
installed in the Keck Telescope.

% -----------------------------------------------------------------------------
\acknowledgements
We thank the Hubble Deep Field team, led by Bob Williams, for
planning, taking, reducing, and making public the phenomenal images of
the HDF.  In addition, Richard Hook, Andy Fruchter and Harry Ferguson
gave us personal assistance with the HST images and catalogs.  We
benefited from helpful conversations with Roger Blandford, Judy Cohen,
George Djorgovski, James Larkin and Chuck Steidel and we thank the
referee, David Koo, for very useful comments.  We are grateful to the
W. M. Keck Foundation, and particularly its president, Howard Keck,
for the vision to fund the construction of the W. M. Keck Observatory.
Financial support was provided by NASA through Grant
No.~AR-06337.13-94A from STScI, which is operated by AURA under NASA
contract NAS5-26555; by the NSF through grants AST-9529170 (DWH) and
AST-9157412 (MAP); and by the Bressler Foundation (MAP).

% -----------------------------------------------------------------------------

% -----------------------------------------------------------------------------
\newpage

\figcaption[Hogg.fig1.ps]{
A diagram showing the locations of NIRC-HDF-A and NIRC-HDF-B with
solid outlines superimposed on the HST/WFPC2 $I_{814}$-band image of
the HDF.  The $I_{814}$-band image is roughly 150~arcsec in width.}

\figcaption[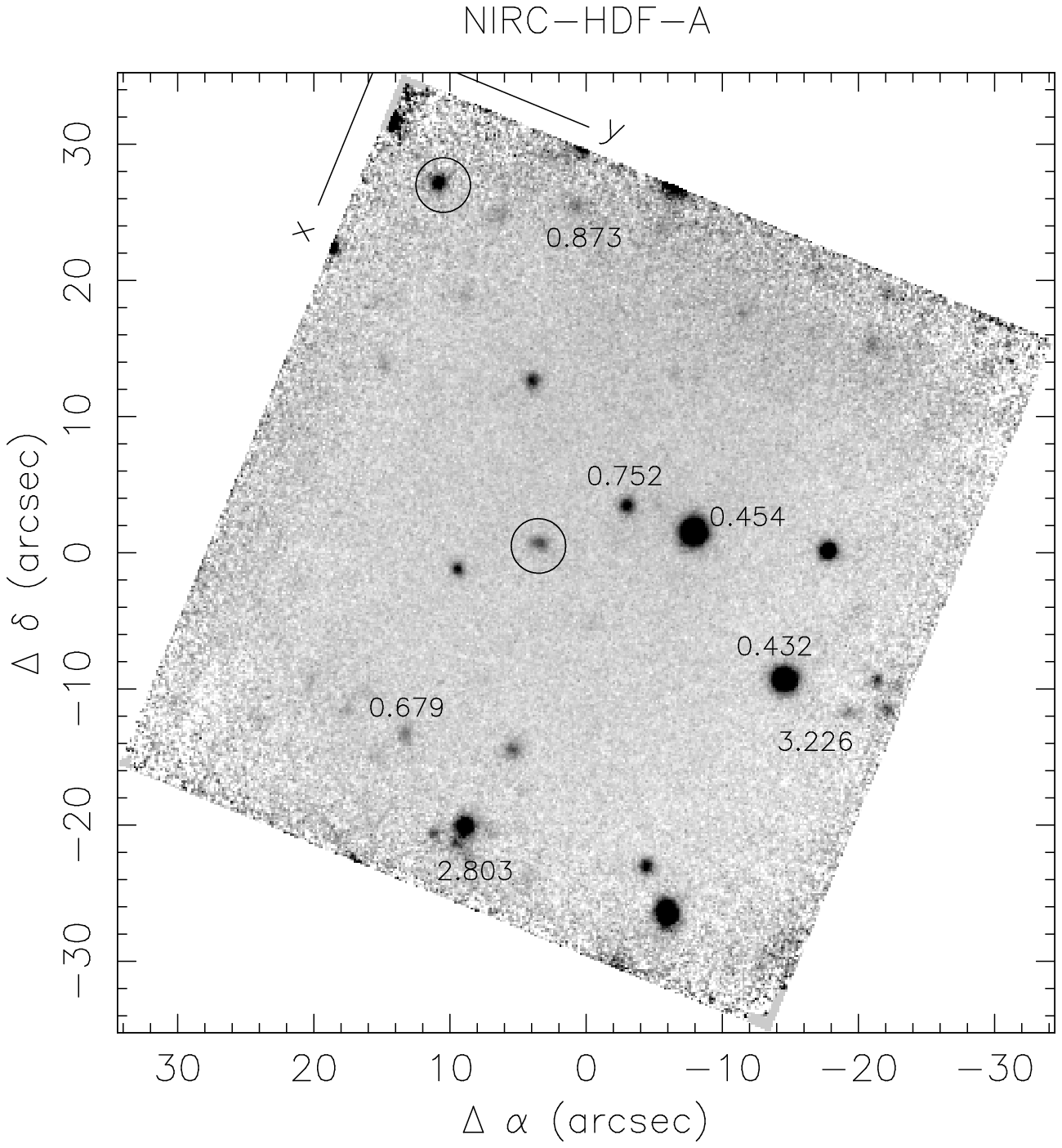]{
The $K$-band image of NIRC-HDF-A with north up.  The mosaic has field
center RA~$12\,36\,44.16$ Dec~$62\,12\,14.9$ (J2000) and position
angle 258~deg.  The objects with $I_{814}-K>4$~mag are circled.
Redshifts from the spectroscopic surveys of Steidel et al (1996) and
Cohen et al (1996) are marked.  The $z=3.226$ source is the very faint
source at $\Delta\alpha,\Delta\delta=-19,-12$ and the source at
$z=2.803$ is not the bright elliptical but the faint, elongated source
at $\Delta\alpha,\Delta\delta=9,-22$.}

\figcaption[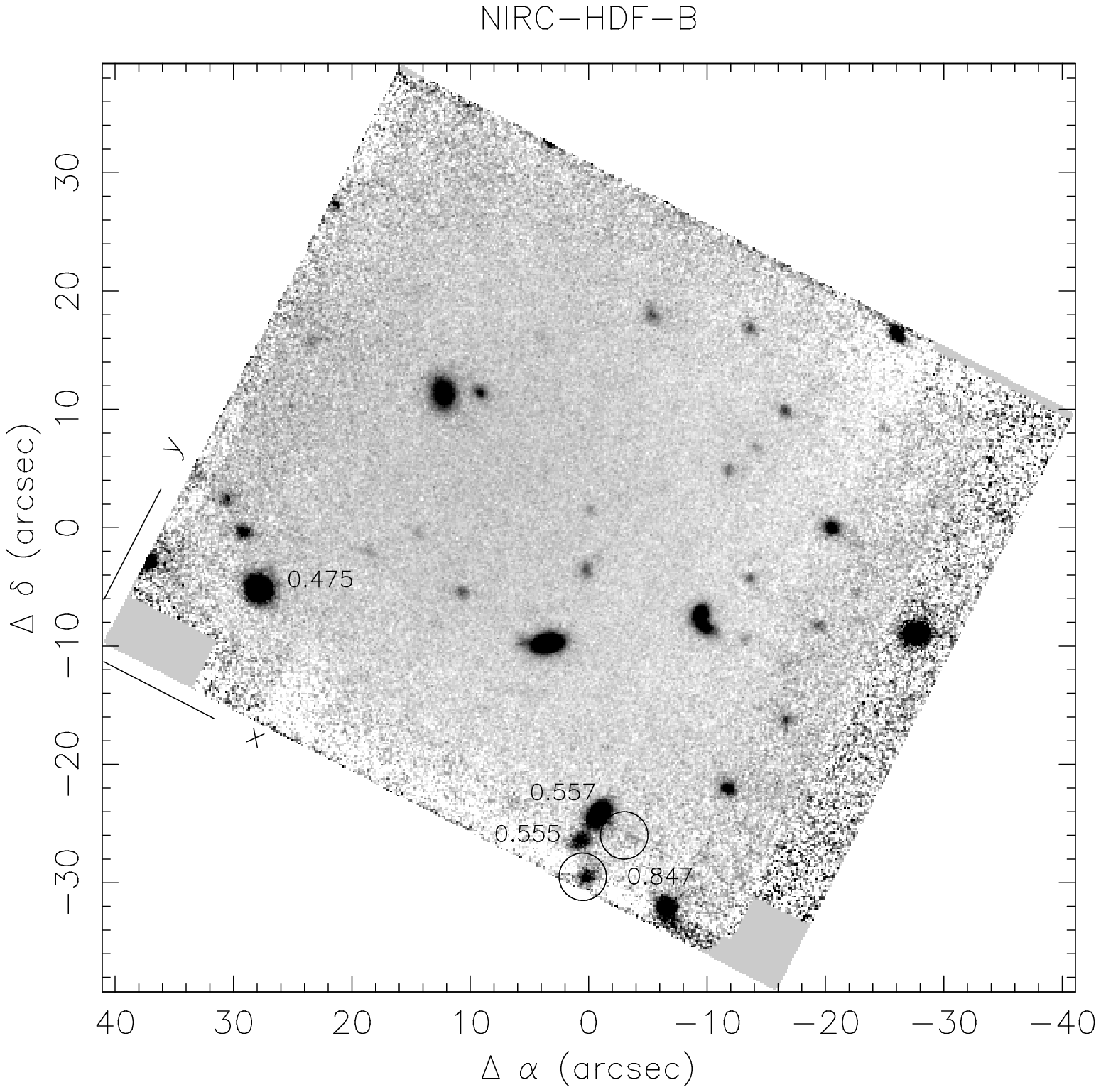]{
The $K$-band image of NIRC-HDF-B with north up.  The mosaic has field
center RA~$12\,36\,44.20$, Dec~$62\,13\,14.4$ (J2000), and position
angle 333~deg.  The objects with $I_{814}-K>4$~mag are circled.
Redshifts from the spectroscopic survey of Cohen et al (1996) are
marked.}

\figcaption[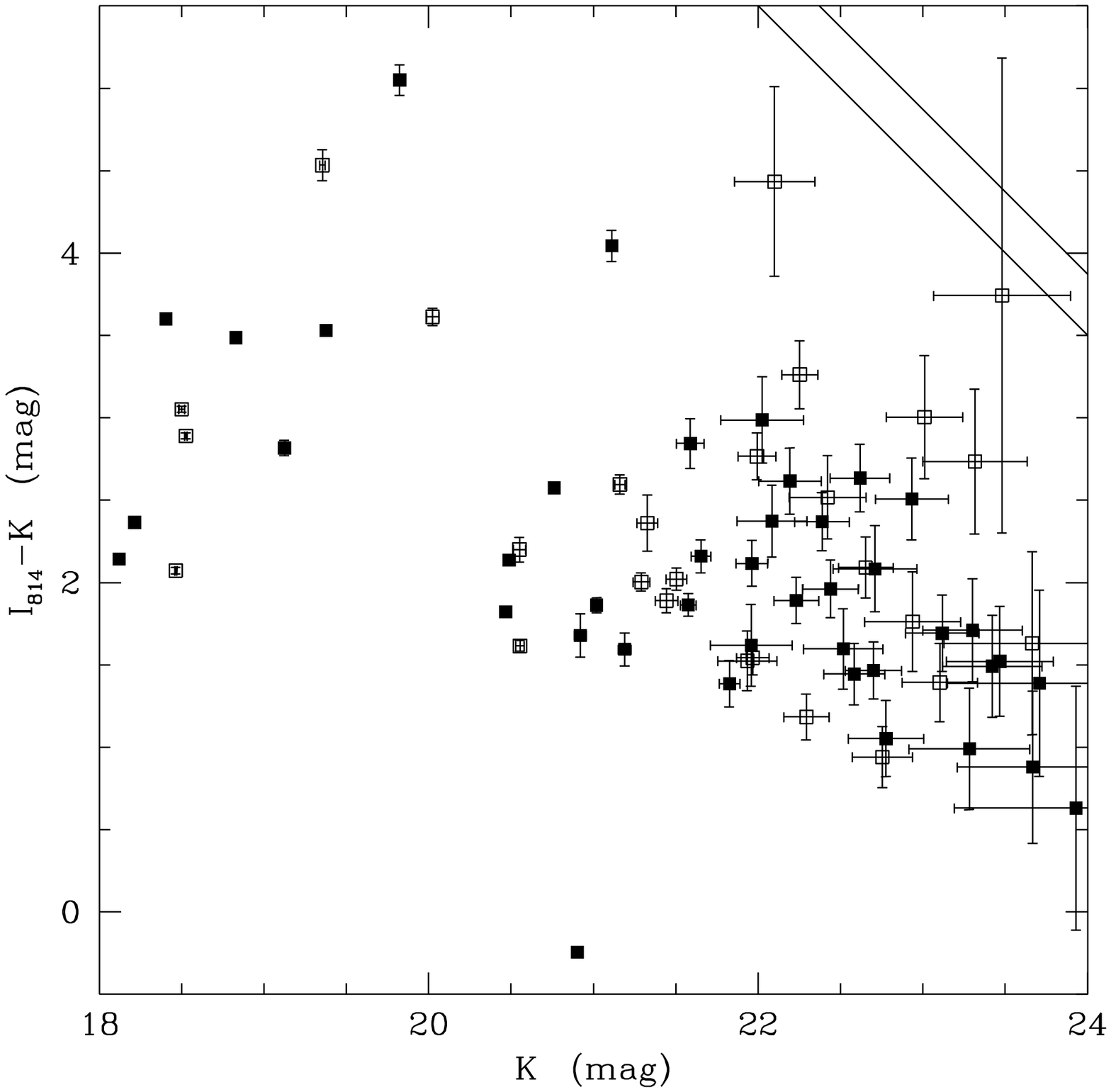]{
The color-magnitude diagram for the $K$-selected sample.  Objects in
NIRC-HDF-A are shown with solid squares and those in NIRC-HDF-B as
open.  All magnitudes are zero-pointed to Vega.  Estimates of the
$1\,\sigma$ detection limits in the smoothed $I_{814}$ images,
27.87~mag for NIRC-HDF-A and 27.50~mag in NIRC-HDF-B, are shown as
diagonal lines, with NIRC-HDF-B at lower sensitivity because it is
centered on the HST/PC1 image.  Objects detected in NIRC-HDF-B but off
the WFPC2 image are not shown.}

\figcaption[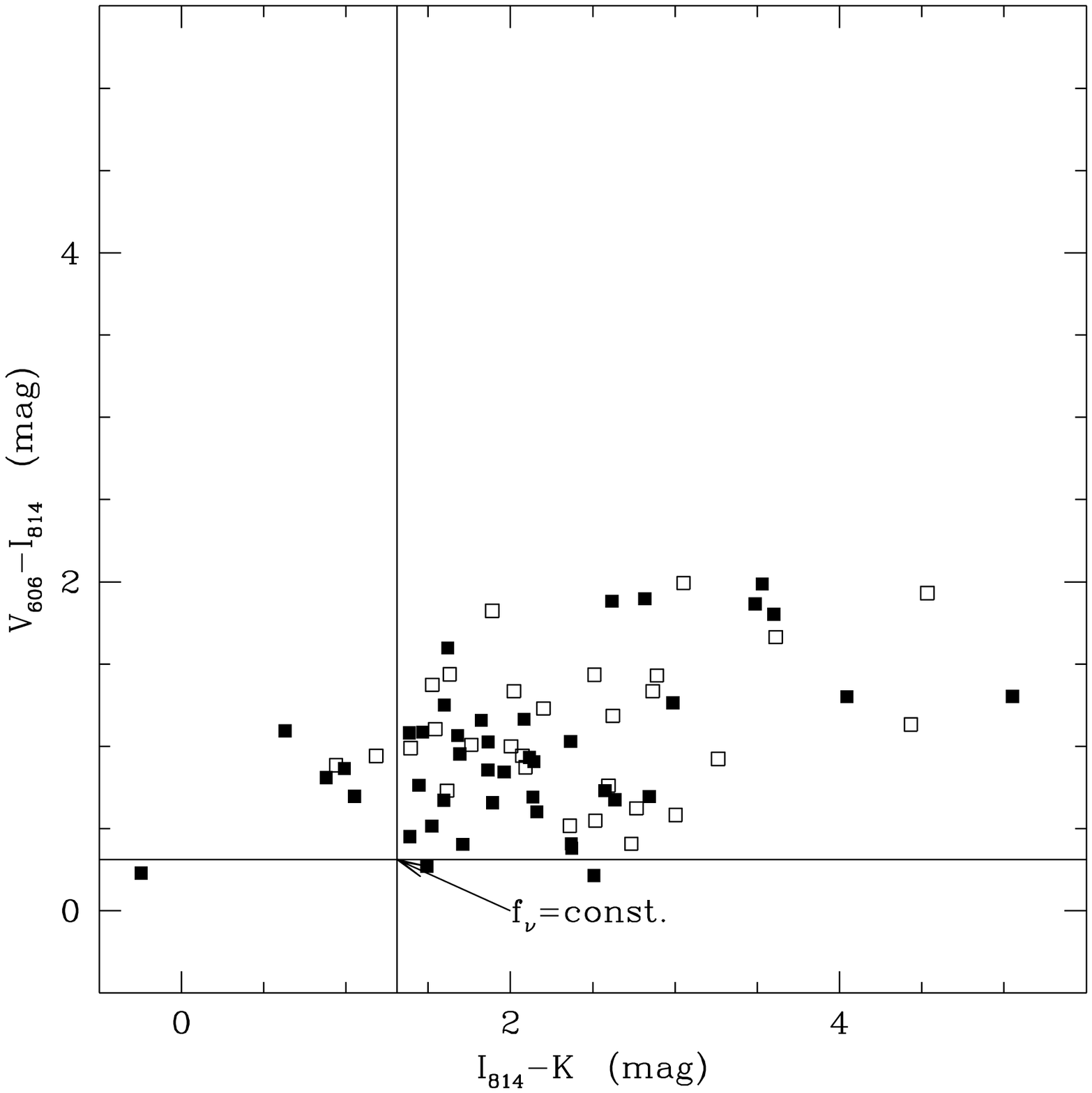]{
The color-color diagram for the $K$-selected sample.  Objects in
NIRC-HDF-A are shown with solid squares and those in NIRC-HDF-B as
open.  Error bars have been dropped to ease confusion; see Tables~1
and 2 for the uncertainties.  All magnitudes are zero-pointed to Vega.
Objects detected in NIRC-HDF-B but off the WFPC2 image are not shown.}

\figcaption[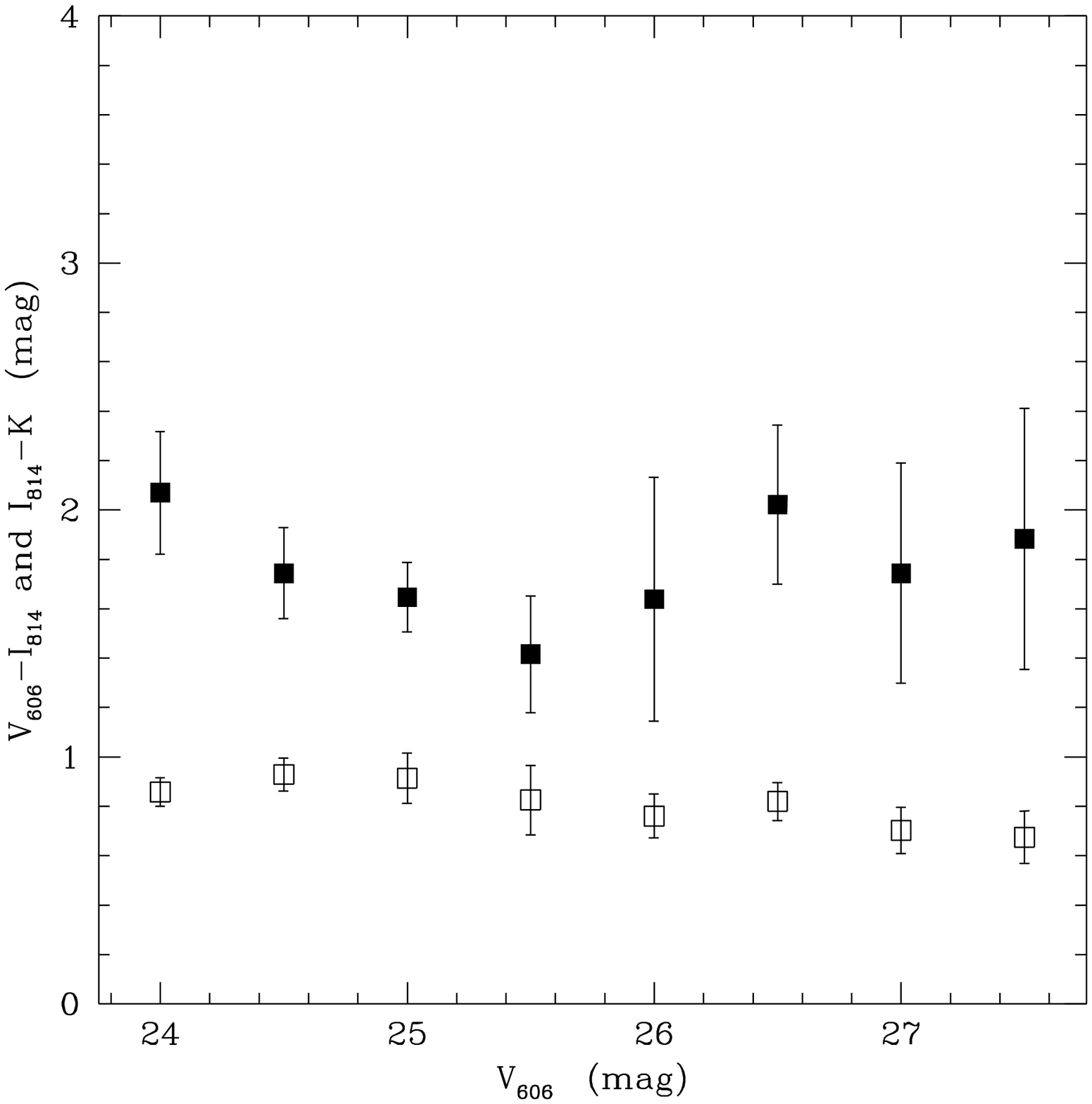]{
Mean color-magnitude diagram for the visually selected sample measured
by averaging together groups of faint objects in one-magnitude-wide
bins (see text).  Open squares mark $V_{606}-I_{814}$ colors and solid
squares mark $I_{814}-K$ colors.  All magnitudes are zero-pointed to
Vega.  The error bars show the scatter computed by bootstrap
resampling.}

\figcaption[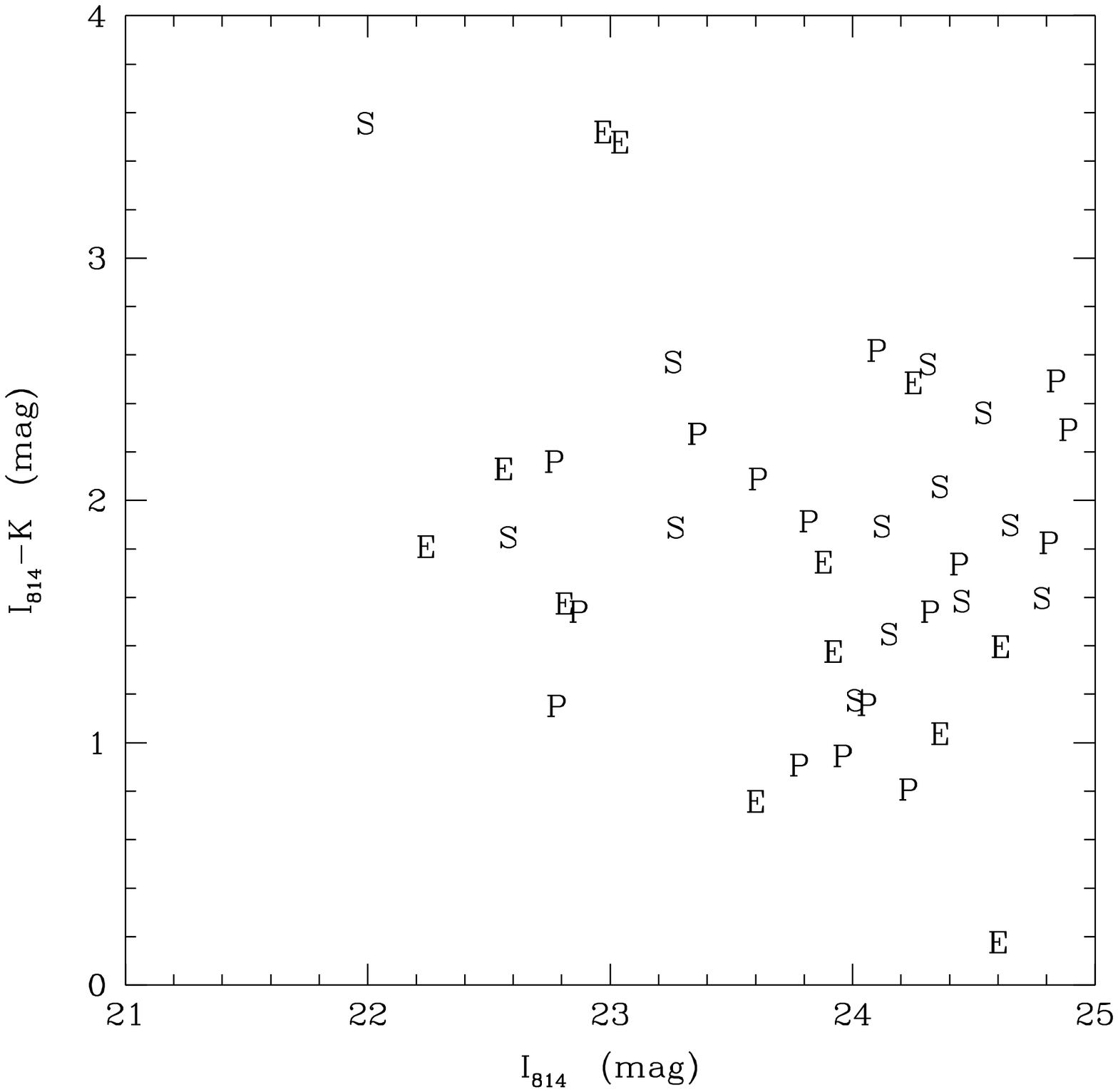]{
Color-magnitude diagram for sources in NIRC-HDF-A classified by
van~den~Bergh et al (1996).  The categories are elliptical (E), spiral
(S) and peculiar (P).  All magnitudes are zero-pointed to Vega.}

% -----------------------------------------------------------------------------
\setcounter{figure}{0}

\begin{figure}
\plotone{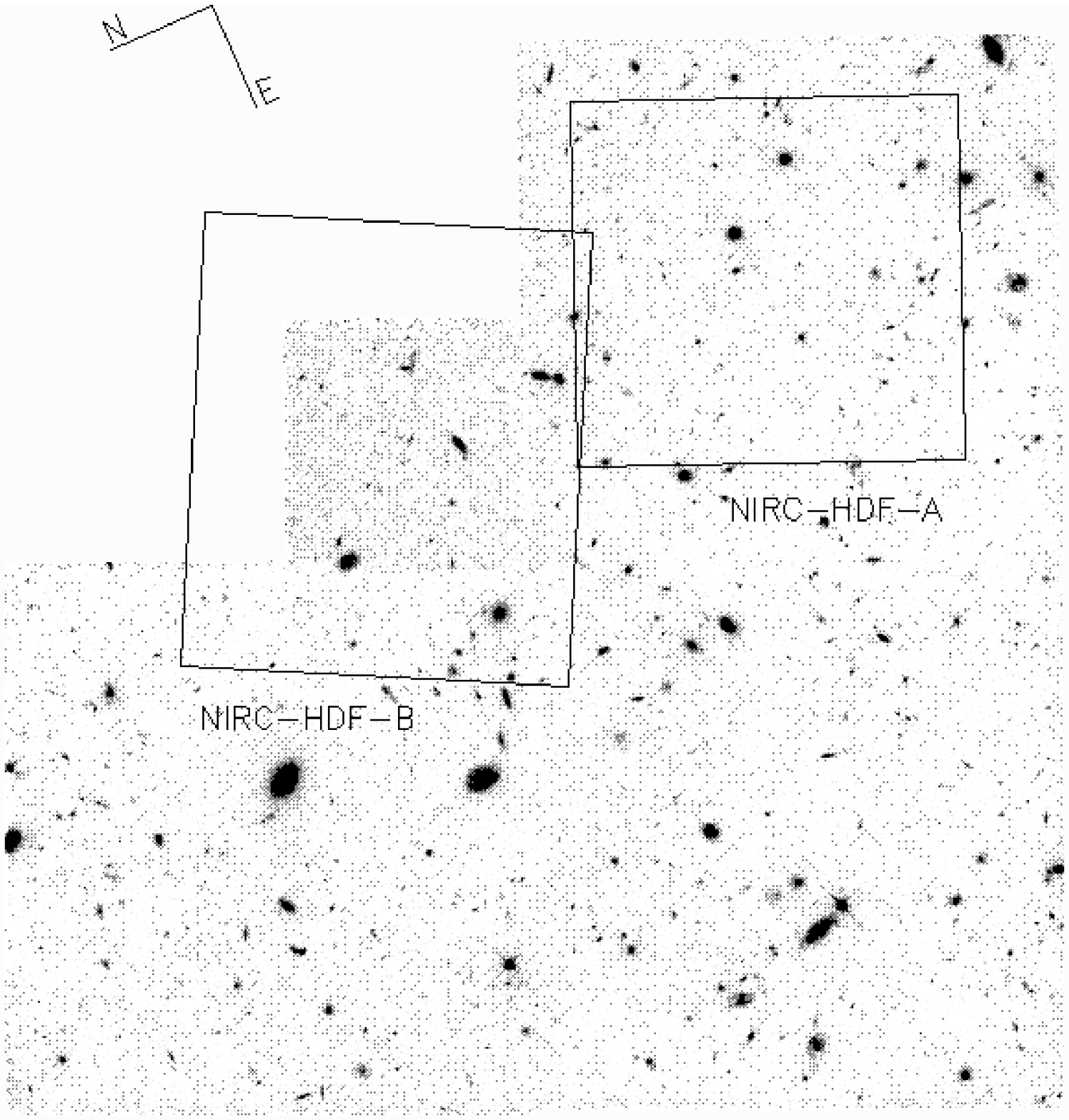}
\caption{~}
\end{figure}

\begin{figure}
\plotone{Hogg.fig2.ps}
\caption{~}
\end{figure}

\begin{figure}
\plotone{Hogg.fig3.ps}
\caption{~}
\end{figure}

\begin{figure}
\plotone{Hogg.fig4.ps}
\caption{~}
\end{figure}

\begin{figure}
\plotone{Hogg.fig5.ps}
\caption{~}
\end{figure}

\begin{figure}
\plotone{Hogg.fig6.ps}
\caption{~}
\end{figure}

\begin{figure}
\plotone{Hogg.fig7.ps}
\caption{~}
\end{figure}

% -----------------------------------------------------------------------------
\clearpage

\begin{deluxetable}{llrrcccl}
\scriptsize
\tablewidth{0pt}
\tablecaption{$K$-selected sources in NIRC-HDF-A}
\tablehead{
   \colhead{RA\tablenotemark{a}}
 & \colhead{Dec.}
 & \colhead{$x$\tablenotemark{b}}
 & \colhead{$y$}
 & \colhead{$K$}
 & \colhead{$V_{606}-I_{814}$\tablenotemark{c}}
 & \colhead{$I_{814}-K$\tablenotemark{c}}
 & \colhead{comments\tablenotemark{d}}
\\
   \colhead{(h\,m\,s)}
 & \colhead{($^{\circ}\,'\,''$)}
 & \colhead{(pixels)}
 & \colhead{(pixels)}
 & \colhead{(mag)}
 & \colhead{(mag)}
 & \colhead{(mag)}
 &
}
\startdata
$12\,36\,46.08$ & $62\,12\,46.9$ & $24$ & $7$ & $19.12\pm 0.01$ & $1.90\pm 0.01$ & $2.82\pm 0.05$ \\
$12\,36\,46.28$ & $62\,12\,33.6$ & $108$ & $32$ & $22.71\pm 0.25$ & $1.16\pm 0.10$ & $2.08\pm 0.26$ & S \\
$12\,36\,45.63$ & $62\,12\,42.4$ & $43$ & $37$ & $19.82\pm 0.02$ & $1.30\pm 0.15$ & $5.05\pm 0.09$ \\
$12\,36\,47.12$ & $62\,12\,12.6$ & $250$ & $48$ & $23.28\pm 0.37$ & $0.86\pm 0.05$ & $0.99\pm 0.37$ & P \\
$12\,36\,46.20$ & $62\,12\,29.0$ & $135$ & $47$ & $22.23\pm 0.14$ & $0.66\pm 0.04$ & $1.89\pm 0.14$ & E \\
$12\,36\,47.54$ & $62\,12\,02.4$ & $318$ & $56$ & $22.52\pm 0.24$ & $1.25\pm 0.05$ & $1.60\pm 0.24$ & P \\
$12\,36\,45.34$ & $62\,12\,34.0$ & $89$ & $70$ & $22.62\pm 0.18$ & $0.68\pm 0.12$ & $2.63\pm 0.20$ \\
$12\,36\,46.98$ & $62\,12\,05.3$ & $291$ & $72$ & $22.78\pm 0.23$ & $0.70\pm 0.03$ & $1.05\pm 0.23$ & E \\
$12\,36\,44.98$ & $62\,12\,40.0$ & $47$ & $71$ & $21.96\pm 0.10$ & $0.93\pm 0.05$ & $2.12\pm 0.14$ & P \\
$12\,36\,46.60$ & $62\,12\,03.2$ & $298$ & $94$ & $22.58\pm 0.18$ & $0.76\pm 0.04$ & $1.44\pm 0.19$ & S \\
$12\,36\,44.16$ & $62\,12\,40.6$ & $29$ & $104$ & $21.83\pm 0.06$ & $1.08\pm 0.02$ & $1.39\pm 0.14$ & P, 0.873 \\
$12\,36\,46.70$ & $62\,11\,53.7$ & $357$ & $113$ & $22.02\pm 0.25$ & $1.26\pm 0.13$ & $2.99\pm 0.26$ \\
$12\,36\,46.28$ & $62\,11\,59.7$ & $313$ & $116$ & $22.44\pm 0.17$ & $0.84\pm 0.06$ & $1.96\pm 0.17$ & S \\
$12\,36\,44.63$ & $62\,12\,27.6$ & $116$ & $116$ & $20.76\pm 0.02$ & $0.73\pm 0.02$ & $2.58\pm 0.03$ & S \\
$12\,36\,43.90$ & $62\,12\,37.0$ & $46$ & $124$ & $22.80\pm 0.25$ & --- & --- & spurious? \\
$12\,36\,45.42$ & $62\,12\,13.6$ & $214$ & $118$ & $20.90\pm 0.03$ & $0.23\pm 0.00$ & $-0.24\pm 0.03$ & stellar \\
$12\,36\,45.97$ & $62\,12\,01.3$ & $298$ & $125$ & $21.58\pm 0.05$ & $1.03\pm 0.03$ & $1.86\pm 0.07$ & S, 0.679 \\
$12\,36\,43.25$ & $62\,12\,39.0$ & $23$ & $147$ & $22.08\pm 0.21$ & $0.38\pm 0.05$ & $2.37\pm 0.22$ & P \\
$12\,36\,45.14$ & $62\,12\,05.4$ & $259$ & $150$ & $23.42\pm 0.30$ & $0.27\pm 0.08$ & $1.49\pm 0.31$ & P \\
$12\,36\,44.55$ & $62\,12\,15.5$ & $188$ & $150$ & $21.11\pm 0.03$ & $1.30\pm 0.15$ & $4.04\pm 0.09$ \\
$12\,36\,43.10$ & $62\,12\,28.2$ & $86$ & $180$ & $22.70\pm 0.17$ & $1.09\pm 0.05$ & $1.47\pm 0.17$ & S \\
$12\,36\,44.83$ & $62\,12\,00.2$ & $285$ & $176$ & $21.02\pm 0.03$ & $0.85\pm 0.02$ & $1.86\pm 0.05$ & S \\
$12\,36\,45.36$ & $62\,11\,54.1$ & $331$ & $169$ & $18.83\pm 0.01$ & $1.86\pm 0.03$ & $3.49\pm 0.02$ & E \\
$12\,36\,43.72$ & $62\,12\,15.8$ & $171$ & $184$ & $23.12\pm 0.22$ & $0.95\pm 0.09$ & $1.69\pm 0.23$ & S \\
$12\,36\,44.73$ & $62\,11\,57.1$ & $302$ & $188$ & $23.67\pm 0.46$ & $0.81\pm 0.06$ & $0.88\pm 0.46$ & E \\
$12\,36\,43.61$ & $62\,12\,18.3$ & $154$ & $183$ & $20.47\pm 0.02$ & $1.16\pm 0.01$ & $1.82\pm 0.02$ & E, 0.752 \\
$12\,36\,42.87$ & $62\,12\,27.9$ & $83$ & $190$ & $23.30\pm 0.30$ & $0.41\pm 0.09$ & $1.71\pm 0.31$ \\
$12\,36\,43.99$ & $62\,12\,09.6$ & $214$ & $188$ & $23.47\pm 0.33$ & $0.51\pm 0.09$ & $1.52\pm 0.33$ \\
$12\,36\,42.37$ & $62\,12\,32.7$ & $46$ & $199$ & $22.39\pm 0.17$ & $0.41\pm 0.07$ & $2.37\pm 0.18$ & E \\
$12\,36\,44.67$ & $62\,11\,50.5$ & $341$ & $207$ & $22.19\pm 0.19$ & $1.88\pm 0.16$ & $2.62\pm 0.20$ & P \\
$12\,36\,42.91$ & $62\,12\,16.4$ & $154$ & $217$ & $18.12\pm 0.00$ & $0.91\pm 0.00$ & $2.14\pm 0.01$ & 0.454 \\
$12\,36\,41.91$ & $62\,12\,26.6$ & $75$ & $234$ & $22.93\pm 0.22$ & $0.21\pm 0.13$ & $2.51\pm 0.25$ \\
$12\,36\,41.64$ & $62\,12\,31.2$ & $42$ & $234$ & $23.71\pm 0.56$ & $0.45\pm 0.10$ & $1.39\pm 0.56$ \\
$12\,36\,43.41$ & $62\,11\,51.4$ & $313$ & $258$ & $20.49\pm 0.03$ & $0.69\pm 0.01$ & $2.14\pm 0.03$ & E \\
$12\,36\,40.84$ & $62\,12\,34.1$ & $10$ & $261$ & $21.96\pm 0.25$ & $1.60\pm 0.04$ & $1.62\pm 0.25$ \\
$12\,36\,40.98$ & $62\,12\,30.2$ & $37$ & $264$ & $21.59\pm 0.08$ & $0.69\pm 0.07$ & $2.84\pm 0.15$ & S \\
$12\,36\,43.19$ & $62\,11\,48.0$ & $330$ & $276$ & $18.40\pm 0.01$ & $1.80\pm 0.02$ & $3.60\pm 0.01$ & S \\
$12\,36\,41.47$ & $62\,12\,15.0$ & $137$ & $281$ & $19.38\pm 0.01$ & $1.99\pm 0.03$ & $3.53\pm 0.01$ & E \\
$12\,36\,41.94$ & $62\,12\,05.4$ & $203$ & $285$ & $18.22\pm 0.01$ & $1.03\pm 0.00$ & $2.37\pm 0.01$ & 0.432 \\
$12\,36\,41.16$ & $62\,12\,10.5$ & $159$ & $306$ & $23.93\pm 0.74$ & $1.09\pm 0.07$ & $0.63\pm 0.74$ & S \\
$12\,36\,41.25$ & $62\,12\,02.9$ & $207$ & $321$ & $21.65\pm 0.06$ & $0.60\pm 0.03$ & $2.16\pm 0.10$ & P, 3.226 \\
$12\,36\,41.91$ & $62\,11\,45.8$ & $321$ & $335$ & $21.34\pm 0.19$ & --- & --- & spurious? \\
$12\,36\,40.91$ & $62\,12\,05.2$ & $187$ & $329$ & $21.19\pm 0.04$ & $0.67\pm 0.01$ & $1.59\pm 0.10$ & E \\
$12\,36\,40.85$ & $62\,12\,03.1$ & $198$ & $337$ & $20.92\pm 0.03$ & $1.07\pm 0.02$ & $1.68\pm 0.13$ \\
\enddata
\tablenotetext{a}{
Positions, in J2000, are found by matching the K-band images to the
astrometric solution given by the HDF team (Williams et al 1996).}
\tablenotetext{b}{
$x$--$y$ positions on final $K$-band mosaics, pixel scale
approximately 0.15~arcsec per pixel.}
\tablenotetext{c}{
$K$-band magnitudes are $1.5$~arcsec diameter aperture magnitudes plus
an aperture correction of $-0.10$~mag.  Colors are measured in
$1.5$~arcsec diameter apertures in the $K$-band and smoothed-$I$-band
images (see text).}
\tablenotetext{d}{
Objects which appear to be spurious noise fluctuations in the $K$-band
image indicated by ``spurious?'' note.  Morphology from van~den~Bergh
et al (1996) indicated by ``E'', ``S'', or ``P''.  Numbers are
spectroscopic redshifts from Cohen et al (1996) or Steidel et al
(1996), when known.  The object marked as $z=2.803$ in Fig.~2 is not
listed because the detection algorithm blends it with the adjacent
bright source.}
\end{deluxetable}

\begin{deluxetable}{llrrcccl}
\scriptsize
\tablewidth{0pt}
\tablecaption{$K$-selected sources in NIRC-HDF-B}
\tablehead{
   \colhead{RA\tablenotemark{a}}
 & \colhead{Dec.}
 & \colhead{$x$\tablenotemark{b}}
 & \colhead{$y$}
 & \colhead{$K$}
 & \colhead{$V_{606}-I_{814}$\tablenotemark{c}}
 & \colhead{$I_{814}-K$\tablenotemark{c}}
 & \colhead{comments\tablenotemark{d}}
\\
   \colhead{(h\,m\,s)}
 & \colhead{($^{\circ}\,'\,''$)}
 & \colhead{(pixels)}
 & \colhead{(pixels)}
 & \colhead{(mag)}
 & \colhead{(mag)}
 & \colhead{(mag)}
 &
}
\startdata
$12\,36\,44.25$ & $62\,12\,52.1$ & $275$ & $49$ & $22.42\pm 0.23$ & $0.55\pm 0.12$ & $2.52\pm 0.25$ \\
$12\,36\,43.13$ & $62\,12\,42.1$ & $351$ & $14$ & $18.50\pm 0.02$ & $1.99\pm 0.01$ & $3.05\pm 0.02$ & 0.847 \\
$12\,36\,45.01$ & $62\,12\,51.1$ & $247$ & $28$ & $21.33\pm 0.06$ & $0.52\pm 0.05$ & $2.36\pm 0.17$ \\
$12\,36\,43.56$ & $62\,12\,48.0$ & $315$ & $40$ & $22.10\pm 0.24$ & $1.13\pm 0.84$ & $4.43\pm 0.58$ & spurious? \\
$12\,36\,44.11$ & $62\,12\,44.8$ & $303$ & $10$ & $19.35\pm 0.02$ & $1.93\pm 0.18$ & $4.53\pm 0.10$ \\
$12\,36\,43.98$ & $62\,12\,49.7$ & $293$ & $41$ & $17.56\pm 0.01$ & $1.19\pm 0.00$ & $2.62\pm 0.01$ & 0.557 \\
$12\,36\,48.10$ & $62\,13\,09.3$ & $65$ & $68$ & $17.40\pm 0.01$ & $1.44\pm 0.00$ & $2.51\pm 0.01$ & 0.475 \\
$12\,36\,44.85$ & $62\,13\,00.4$ & $225$ & $85$ & $23.48\pm 0.41$ & --- & $3.74\pm 1.44$ \\
$12\,36\,42.39$ & $62\,12\,52.3$ & $350$ & $90$ & $19.66\pm 0.03$ & --- & --- & off-WFPC2 \\
$12\,36\,48.30$ & $62\,13\,14.1$ & $43$ & $93$ & $20.02\pm 0.04$ & $1.66\pm 0.08$ & $3.61\pm 0.05$ \\
$12\,36\,45.95$ & $62\,13\,06.3$ & $163$ & $97$ & $23.01\pm 0.23$ & $0.58\pm 0.35$ & $3.00\pm 0.37$ \\
$12\,36\,48.50$ & $62\,13\,17.0$ & $26$ & $105$ & $20.55\pm 0.04$ & $1.23\pm 0.03$ & $2.20\pm 0.08$ \\
$12\,36\,46.76$ & $62\,13\,12.3$ & $111$ & $115$ & $22.29\pm 0.14$ & $0.94\pm 0.04$ & $1.18\pm 0.14$ \\
$12\,36\,45.65$ & $62\,13\,08.9$ & $167$ & $118$ & $21.29\pm 0.05$ & $1.00\pm 0.03$ & $2.00\pm 0.06$ \\
$12\,36\,44.59$ & $62\,13\,04.6$ & $223$ & $115$ & $17.81\pm 0.00$ & $1.33\pm 0.01$ & $2.87\pm 0.01$ \\
$12\,36\,46.17$ & $62\,13\,14.0$ & $130$ & $137$ & $22.75\pm 0.18$ & $0.88\pm 0.04$ & $0.94\pm 0.19$ \\
$12\,36\,41.68$ & $62\,12\,58.1$ & $362$ & $139$ & $20.84\pm 0.06$ & --- & --- & off-WFPC2 \\
$12\,36\,42.18$ & $62\,13\,04.9$ & $321$ & $168$ & $21.99\pm 0.12$ & $0.62\pm 0.10$ & $2.77\pm 0.14$ \\
$12\,36\,41.65$ & $62\,13\,02.9$ & $349$ & $168$ & $22.26\pm 0.20$ & --- & --- & off-WFPC2 \\
$12\,36\,44.10$ & $62\,13\,10.9$ & $224$ & $163$ & $21.16\pm 0.03$ & $0.76\pm 0.04$ & $2.60\pm 0.06$ \\
$12\,36\,42.70$ & $62\,13\,06.7$ & $294$ & $167$ & $18.53\pm 0.01$ & $1.43\pm 0.01$ & $2.89\pm 0.02$ \\
$12\,36\,44.07$ & $62\,13\,16.0$ & $210$ & $193$ & $22.25\pm 0.11$ & $0.92\pm 0.25$ & $3.26\pm 0.21$ \\
$12\,36\,41.28$ & $62\,13\,06.1$ & $354$ & $194$ & $21.15\pm 0.07$ & --- & --- & off-WFPC2 \\
$12\,36\,44.62$ & $62\,13\,19.1$ & $178$ & $199$ & $23.10\pm 0.23$ & $0.99\pm 0.09$ & $1.39\pm 0.24$ \\
$12\,36\,42.10$ & $62\,13\,10.1$ & $308$ & $200$ & $21.44\pm 0.07$ & $1.82\pm 0.05$ & $1.89\pm 0.07$ \\
$12\,36\,47.44$ & $62\,13\,30.5$ & $28$ & $207$ & $21.93\pm 0.18$ & $1.37\pm 0.04$ & $1.52\pm 0.18$ \\
$12\,36\,42.04$ & $62\,13\,12.0$ & $305$ & $213$ & $23.66\pm 0.54$ & $1.44\pm 0.27$ & $1.63\pm 0.56$ \\
$12\,36\,40.81$ & $62\,13\,08.9$ & $365$ & $220$ & $22.16\pm 0.22$ & --- & --- & off-WFPC2 \\
$12\,36\,43.17$ & $62\,13\,17.9$ & $241$ & $223$ & $23.32\pm 0.32$ & $0.41\pm 0.35$ & $2.73\pm 0.44$ \\
$12\,36\,40.09$ & $62\,13\,05.4$ & $405$ & $215$ & $17.02\pm 0.00$ & --- & --- & off-WFPC2 \\
$12\,36\,45.86$ & $62\,13\,25.9$ & $107$ & $213$ & $18.46\pm 0.01$ & $0.94\pm 0.01$ & $2.07\pm 0.02$ \\
$12\,36\,45.41$ & $62\,13\,26.0$ & $125$ & $223$ & $20.55\pm 0.03$ & $0.73\pm 0.01$ & $1.61\pm 0.03$ \\
$12\,36\,41.10$ & $62\,13\,14.4$ & $336$ & $246$ & $19.55\pm 0.02$ & --- & --- & off-WFPC2 \\
$12\,36\,42.34$ & $62\,13\,19.3$ & $271$ & $249$ & $21.50\pm 0.06$ & $1.33\pm 0.05$ & $2.02\pm 0.07$ \\
$12\,36\,46.16$ & $62\,13\,34.7$ & $68$ & $258$ & $22.94\pm 0.29$ & $1.01\pm 0.12$ & $1.76\pm 0.30$ \\
$12\,36\,44.64$ & $62\,13\,30.5$ & $143$ & $266$ & $22.65\pm 0.17$ & $0.87\pm 0.11$ & $2.09\pm 0.19$ \\
$12\,36\,42.04$ & $62\,13\,21.3$ & $277$ & $267$ & $21.97\pm 0.10$ & $1.10\pm 0.04$ & $1.54\pm 0.10$ \\
$12\,36\,39.86$ & $62\,13\,16.6$ & $380$ & $285$ & $22.29\pm 0.33$ & --- & --- & off-WFPC2 \\
$12\,36\,41.67$ & $62\,13\,24.4$ & $283$ & $292$ & $21.04\pm 0.04$ & --- & --- & off-WFPC2 \\
$12\,36\,41.28$ & $62\,13\,23.6$ & $301$ & $296$ & $23.91\pm 0.82$ & --- & --- & off-WFPC2 \\
$12\,36\,40.47$ & $62\,13\,22.9$ & $336$ & $309$ & $22.36\pm 0.26$ & --- & --- & off-WFPC2 \\
$12\,36\,43.29$ & $62\,13\,32.6$ & $192$ & $307$ & $20.87\pm 0.04$ & --- & --- & off-WFPC2 \\
$12\,36\,39.17$ & $62\,13\,20.3$ & $397$ & $321$ & $20.62\pm 0.17$ & --- & --- & off-WFPC2 \\
$12\,36\,42.11$ & $62\,13\,31.4$ & $244$ & $325$ & $21.15\pm 0.05$ & --- & --- & off-WFPC2 \\
\enddata
\tablenotetext{a}{
Positions, in J2000, are found by matching the K-band images to the
astrometric solution given by the HDF team (Williams et al 1996).}
\tablenotetext{b}{
$x$--$y$ positions on final $K$-band mosaics, pixel scale
approximately 0.15~arcsec per pixel.}
\tablenotetext{c}{
$K$-band magnitudes are $1.5$~arcsec diameter aperture magnitudes plus
an aperture correction of $-0.10$~mag.  Colors are measured in
$1.5$~arcsec diameter apertures in the $K$-band and smoothed-$I$-band
images (see text).}
\tablenotetext{d}{
Objects off the WFPC2 field indicated by ``off-WFPC2'' note.  Numbers
are spectroscopic redshifts from Cohen et al (1996), when known.  The
objects with $z=0.555$ and $0.557$ shown in Fig.~3 are blended into
one by the detection algorithm.}
\end{deluxetable}

\newpage

\begin{references}
\reference{}
Aaronson, M. 1977, PhD Thesis, Harvard University
\reference{}
Abraham, R. G. 1996b, in {\it Proc.\ 37 Herst.\ Conf: HST and the High
Redshift Universe,\/} Tanvir, N. R., Aragon-Salamanca, A. \& Wall,
J. V., eds., in press
\reference{}
Babul, A. \& Ferguson, H. C. 1996, ApJ, 458, 100
\reference{}
Bertin, E. \& Arnouts, S. 1996, A\&AS, in press
\reference{}
Casali, M. M. \& Hawarden, T. G. 1992, JCMT-UKIRT Newsletter, 3, 33
\reference{}
Cohen, J. G., Cowie, L. L., Hogg, D. W., Songaila, A., Blandford, R.,
Hu, E. M. \& Shopbell, P. 1996, ApJ 471 L5
\reference{}
Coleman, G. D., Wu, C.-C. \& Weedman, D. W. 1980, ApJS, 43, 393
\reference{}
Djorgovski, S., Soifer, B. T., Pahre, M. A., Larkin, J., Smith, J. D.,
Neugebauer, G., Smail, I., Matthews, K., Hogg, D. W., Blandford,
R. D., Cohen, J., Harrison, W. \& Nelson, J. 1995, ApJ, 438, L13
\reference{}
Efron, B. \& Tibrishani, R. 1991, Science, 253, 390
\reference{}
Giavalisco, M., Livio, M., Bohlin, R. C., Macchetto, F. D. \& Stecher,
T. P. 1996, AJ, 112, 369
\reference{}
Katz, N. 1992, ApJ, 391, 502
\reference{}
Matthews, K. \& Soifer, B. T. 1994, in {\em Infrared Astronomy with
Arrays: The Next Generation,\/} ed.\ McLean, I. (Dordrecht: Kluwer),
239
\reference{}
O'Connell, R. W. \& Marcum, P. 1996, in {\it Proc.\ 37 Herst.\ Conf:
HST and the High Redshift Universe,\/} Tanvir, N. R.,
Aragon-Salamanca, A. \& Wall, J. V., eds., in press
\reference{}
Pascarelle, S. M., Windhorst, R. A., Keel, W. C. \& Odewahn, S. C.,
1996, Nature, 383, 45
\reference{}
Persson, S. E. 1996, private communication
\reference{}
Steidel, C. C., Giavalisco, M., Dickinson, M. \& Adelberger, K. 1996,
AJ, 112, 352
\reference{}
van den Bergh, S., Abraham, R. G., Ellis, R. S., Tanvir, N. R.,
Santiago, B. X. \& Glazebrook, K. G. 1996, AJ, 112, 359
\reference{}
Williams, R. E. et al. 1996, AJ, 112, 1335
\end{references}
\end{document}